\documentclass[conference]{IEEEtran}
\IEEEoverridecommandlockouts

\usepackage{color}
\usepackage{cite}
\usepackage{tabularx}
\usepackage{tikz}

\usepackage{soul}
\usepackage{multirow}
\usepackage{cite}
 
\usepackage{graphicx}
\usepackage{float}
\usepackage{subfigure}
\usepackage{cite} 
\usepackage{amsmath}

\usepackage{cite}
\usepackage{amsmath,amsfonts}
\usepackage{algorithm}
\usepackage{array}
\usepackage{algpseudocode}
\usepackage[caption=false,font=normalsize,labelfont=sf,textfont=sf]{subfig}
\usepackage{textcomp}
\usepackage{stfloats}
\usepackage{url}
\usepackage{verbatim}
\usepackage{graphicx}
\usepackage{subfigure}
\usepackage{cite}
\usepackage{pifont}
\usepackage{makecell}
\usepackage{bm}
\usepackage{bbm}
\usepackage[english]{babel}
\usepackage{amsthm}
\usepackage{subcaption}
\usepackage{color}
\usepackage{subfigure}

\IEEEoverridecommandlockouts

\usepackage{tabularx}
\usepackage{tikz}
\usepackage{soul}
\usepackage{multirow}
 
\usepackage{graphicx}
\usepackage{float}

\begin{document}

\title{Energy Efficient Federated Learning with Hyperdimensional Computing (HDC)}

\author{\IEEEauthorblockN{Yahao Ding\IEEEauthorrefmark{1}, Yinchao Yang\IEEEauthorrefmark{1}, Jiaxiang Wang\IEEEauthorrefmark{1}, Zhonghao Liu\IEEEauthorrefmark{1},\\
Zhaohui Yang\IEEEauthorrefmark{2}, Mingzhe Chen\IEEEauthorrefmark{3}, Mohammad Shikh-Bahaei\IEEEauthorrefmark{1}}
\IEEEauthorblockA{\IEEEauthorrefmark{1}King's College London,\IEEEauthorrefmark{2} Zhejiang University,\IEEEauthorrefmark{3} University of Miami,\\
Email: \IEEEauthorrefmark{1}{\{yahao.ding, yinchao.yang, jiaxiang.wang, zhonghao.liu, m.sbahaei\}}@kcl.ac.uk,\\
\IEEEauthorrefmark{2}yang\_zhaohui@zju.edu.cn, \IEEEauthorrefmark{3}mingzhe.chen@miami.edu }
}

\markboth{Journal of \LaTeX\ Class Files,~Vol.~14, No.~8, August~2015}%
{Shell \MakeLowercase{\textit{et al.}}: Bare Demo of IEEEtran.cls for IEEE Journals}

\maketitle
\begin{abstract} 
This paper investigates the problem of minimizing total energy consumption for secure federated learning (FL) in wireless edge networks, a key paradigm for decentralized big data analytics. To tackle the high computational cost and privacy challenges of processing large-scale distributed data with conventional neural networks, we propose an FL with hyperdimensional computing and differential privacy (FL-HDC-DP) framework. Each edge device employs hyperdimensional computing (HDC) for lightweight local training and applies differential privacy (DP) noise to protect transmitted model updates. The total energy consumption is minimized through a joint optimization of the HDC dimension, transmit power, and CPU frequency. An efficient hybrid algorithm is developed, combining an outer enumeration search for HDC dimensions with an inner one-dimensional search for resource allocation. Simulation results show that the proposed framework achieves up to 83.3\% energy reduction compared with baseline schemes, while maintaining high accuracy and faster convergence.

\end{abstract}

\begin{IEEEkeywords}
Federated learning, hyperdimensional computing, differential privacy, resource allocation, energy efficiency.
\end{IEEEkeywords}

\IEEEpeerreviewmaketitle


\section{Introduction}

The contemporary shift towards a data-driven digital economy is fueled by big data, which offers immense value in understanding complex systems. An explosive growth in data generation from Internet of Things (IoT) devices, smartphones, and wearable technologies is creating massive data volumes at the network edge. This decentralized data deluge presents new challenges and opportunities, demanding advanced architectures for big data processing that move beyond traditional centralized models. Federated Learning (FL) has emerged as a key paradigm integrating big data with artificial intelligence (AI), supporting collaborative machine learning directly on edge devices. By training models locally and sharing only model updates, FL obviates the need to transfer raw data, representing a transformative approach to privacy protection in big data environments.

Despite its promise, the large-scale deployment of FL for edge-based big data analytics faces two fundamental obstacles. First, the efficient transmission and processing of massive data are constrained by a fundamental energy bottleneck. Edge devices are typically battery-powered, yet the FL process is inherently energy-intensive. This consumption stems from both local computation, especially for complex models like neural networks (NN), and wireless communication required to transmit model updates over communication networks. Second, security and trust remain core concerns. Standard FL updates are vulnerable to inference attacks that can reveal sensitive information about local data, which underscores the need for robust privacy guarantees.

To address the challenge of high computational energy, hyperdimensional computing (HDC) provides a compelling solution as an emerging, brain-inspired data analytic model. Unlike NNs model, HDC represents information in a high-dimensional vector space and relies on computationally lightweight vector operations such as bundling, binding, and permutation. This makes it suitable for resource-constrained edge devices. Concurrently, to resolve privacy leakage risks, differential privacy (DP) has been established as the standard method for providing rigorous, provable privacy guarantees through the addition of controlled noise.

Existing research has explored these challenges from various perspectives. In the domain of resource efficiency, efforts have focused on minimizing energy consumption in traditional FL frameworks. For instance, the work in \cite{9264742} proposed a comprehensive joint optimization problem, collaboratively adjusting transmission time, power, bandwidth, and computation frequency. The authors in \cite{9916128} similarly focused on minimizing total energy by jointly optimizing weight quantization and wireless resource allocation. 
Other works have targeted reducing communication overhead and latency. For example, \cite{hsieh2021fl} and \cite{10473907} demonstrated improved communication efficiency through bipolarized hypervectors and the novel HyperFeel framework, respectively. The work in \cite{10925099} further combined federated split learning with HDC, designing an algorithm to minimize the maximum user transmission time.


Research in privacy and security is equally critical. To defend against gradient leakage attacks, the work in \cite{10660465} has focused on optimizing DP noise and layer selection. HDC introduces its own security considerations, and recent work \cite{khaleghi2024private} has designed frameworks to achieve efficient and secure private learning. More advanced system designs are also emerging. The FedHDPrivacy framework \cite{piran2025privacy} was designed to balance privacy and accuracy, while other studies have focused on robustness in noisy FL scenarios \cite{morris2022} and the hardware-level energy efficiency of HDC processors \cite{9813404}. Recent studies have also conducted systematic analyses of aggregation strategies in HDC-FL \cite{zhang2023hyperdimensional} and explored the fusion of large AI models with HDC in future 6G networks \cite{xu2025new}.

A clear research gap thus remains. Although previous works have separately addressed energy, privacy, or HDC, there is a lack of a unified big data framework capable of jointly modeling and co-optimizing the computational efficiency of HDC, the privacy guarantees of DP, and the system-level computation and communication resources. Such a framework is essential for fundamentally resolving the total energy consumption issue in secure, decentralized big data analytics on wireless edge environments.

To fill this critical gap, this paper proposes a novel framework: FL with HDC and DP (FL-HDC-DP). This framework is specifically designed for energy-constrained wireless edge networks. It enables users to leverage the computational efficiency of HDC for local model training while obtaining robust privacy protection through the mathematical rigor of DP. Our central objective is to minimize the total energy consumption of all participating users via a comprehensive joint optimization strategy. This work aims to make AI-driven big data analysis feasible on large-scale edge devices. The main contributions of this paper are summarized as follows:
\begin{itemize}
\item We propose an FL-HDC-DP framework designed for energy-constrained wireless edge networks. This framework leverages the computational efficiency of HDC for local model training and the mathematical rigor of DP for robust privacy protection.
\item We formulate a joint optimization problem aimed at minimizing the total
energy consumption, encompassing both local computation and wireless communication. This problem is the first to uniformly model and co-optimize the HDC model dimension alongside system resources such as transmission power and CPU frequency. 
\item We design and validate an efficient, low-complexity hybrid algorithm to solve the non-convex problem. The algorithm utilizes an outer enumeration search to explore HDC dimensions and an inner one-dimensional search to efficiently find the optimal resource allocation for each decoupled subproblem. Simulation results demonstrate that this joint optimization approach achieves up to 83.3\% in total energy savings compared to baseline schemes.

\end{itemize}


\vspace{-0.3em}
\section{System Model}
\label{sec:system}

\begin{figure}[t]
\centering\includegraphics[width=0.5\textwidth]{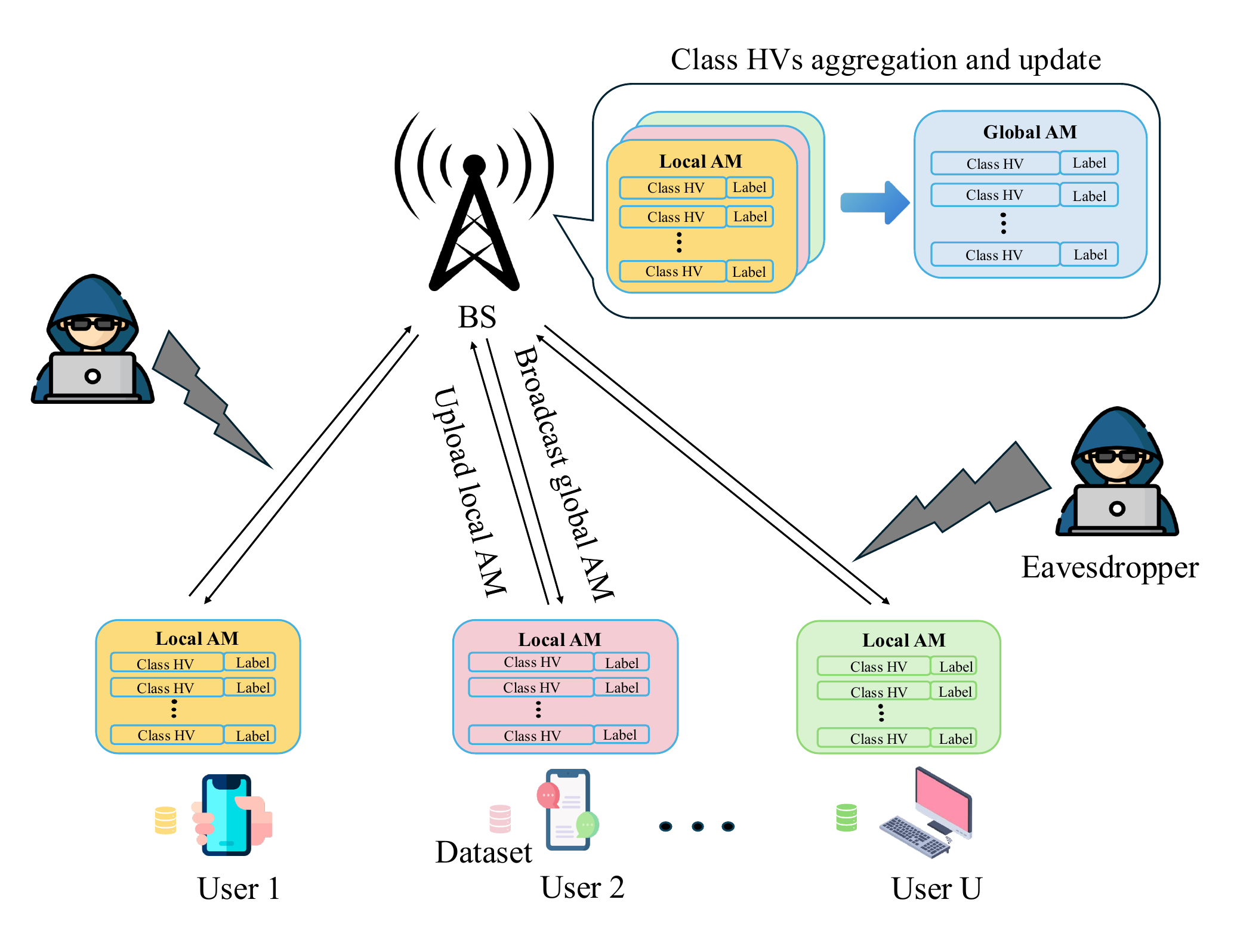}
    \caption{The FL-HDC-DP model over wireless communication networks.}
    \label{fig:FL-HDC-DP}
\end{figure}

\begin{algorithm}[t]
\caption{FL-HDC-DP Algorithm}
\label{alg:fl-hdc-dp}
\begin{algorithmic}[1]
\Statex \textbf{Input:}User set $\mathcal{U}$; class set $\mathcal{N}$; user dataset $\mathcal{S}_i$, round $J$, learning rate $\eta$.
\Statex \textbf{Output:} Final global AM $\boldsymbol{A}_g^J$. 
\For{$j = 0$ to $J-1$}
\State BS broadcasts global AM $\boldsymbol{A}_g^0$ and bipolar HV benchmarks to all users.
\For{user $i \in \mathcal{U}$ in parallel}
        \If{$j=0$}
        \State Encode local data to obtain $\boldsymbol{H}_{i,l}$ and clip them.
        \State Aggregate all HVs to obtain local AM $\boldsymbol{A}_{i,n}$
        \Else
            \State Perform retraining guided by $\boldsymbol{A}_g^j$ to update the local AM according to Eq.\eqref{retrain}, yielding $\boldsymbol{A}_i^{j+1}$.
        \EndIf 
        \State Add DP noise to $\boldsymbol{A}_i$: $\boldsymbol{A}_i^{j+1} \leftarrow \boldsymbol{A}_i^{j+1} + \boldsymbol{n}_i^{j+1}$.
        \State Transmit $\boldsymbol{A}_i^{j+1}$ to the BS.
    \EndFor
    
    \Statex \textbf{Server-side Aggregation:}
    \State Receive $\{\boldsymbol{A}_i^{j+1}\}_{i=1}^U$ from all users.
    \State Conduct element-wise
average to update the global AM: $\boldsymbol{A}_g^{j+1} \leftarrow \frac{1}{U} \sum_{i=1}^U \boldsymbol{A}_i^{j+1}$.
    \State Broadcast $\boldsymbol{A}_g^{j+1}$ to all users for the next round.
\EndFor
\end{algorithmic}
\end{algorithm}

We consider a single-cell network with one base station (BS) serving $U$ users, as shown in Fig.~\ref{fig:FL-HDC-DP}. They cooperatively train a FL model over wireless networks for classification inference. 
To reduce the computational burden on resource-constrained edge devices and to protect user privacy against an honest-but-curious server and external eavesdroppers, we adopt a secure FL framework that combines HDC with DP, named FL-HDC-DP. 
Let $\mathcal{U}=\{1,2,\ldots,U\}$ denote the user set. 
User $i\in \mathcal{U}$ holds a private dataset $\mathcal{S}_i=\{(\boldsymbol{x}_{i\ell},y_{i\ell})\}_{\ell=1}^{S_i}$ with inputs $\boldsymbol{x}_{i\ell}\in\mathbb{R}^{m}$ and labels $y_{i\ell}\in \{1,\ldots,N\}$. 

\subsection{HDC Model}
HDC is an emerging, brain-inspired computing paradigm that operates by representing and manipulating information in a vector space with thousands of dimensions. The fundamental element of HDC is the hypervector (HV), denoted as $\boldsymbol{H}$. A $d$-dimensional HV can be expressed as ${\boldsymbol{H}}=\left\langle h_1, h_2, \ldots, h_d\right\rangle$, where $h_{i}$ represents the value in the 
${i}$-th dimension. Unlike NN, which are predicated on floating-point arithmetic and backpropagation, HDC relies on a set of computationally lightweight vector operations, primarily bundling, binding, and permutation. 
\subsubsection{Encoding}
The encoding process transforms input samples into HVs. We use record-based encoding. First, Item Memories (IMs) provide HVs for feature positions ($\boldsymbol{H}_p$) and feature values ($\boldsymbol{H}_v$). For a sample $\ell$, its position HV and value HV are bound and then bundled with all other features to generate the sample HV $H_\ell$. Using the MNIST dataset (784 pixels) as an example, the encoding process is:
   \begin{equation}
     \boldsymbol{H}_\ell = \sum_{k=0}^{783} \boldsymbol{H}_{\ell,p(k)} \odot \boldsymbol{H}_{\ell,v(\text{pixel} (k))}.
     \label{encoding}
   \end{equation}
where $k$ is the index of the pixel position. $\boldsymbol{H}_{\ell,p(k)}$ and $\boldsymbol{H}_{\ell,v(\text{pixel}(k))}$ represent the position HV corresponding to the $k$-th pixel and the value HV corresponding to the value of the $k$-th pixel, respectively, for the $\ell$-th sample. $\odot$ is the element-wise product.


\subsubsection{Training}
Training involves adding all encoded sample HVs of the same class $n$ to obtain the class HVs $\boldsymbol{A}^n$, which can be obtained by
\begin{equation}
\boldsymbol{A}^n=\sum_{m=1}^{M_n} \boldsymbol{H}_{nm},
\label{A1}
\end{equation}
where $M_n$ is the number of samples in class $n$ and $\boldsymbol{H}_{nm}$ represents the HV corresponding to the $m$-th sample of class $n$. Aggregating all the class HVs to form the associative memory (AM) $\boldsymbol{A}=\left\{{\boldsymbol{A}^1}, {\boldsymbol{A}^2}, \ldots, {\boldsymbol{A}^N}\right\}$.
HDC can achieve reasonable accuracy with just a single training pass.

\subsubsection{Inference}
During inference, an input is encoded into a query HV, denoted as $H_q$, using the same encoding process. This $H_q$ is then compared to every class HV $\boldsymbol{A}^n$ using a similarity measure. We use cosine similarity:
\begin{equation}
\begin{array}{r}
\xi_{\text {cosine }}\left(\boldsymbol{H}_q, \boldsymbol{A}^n\right)=\frac{{\boldsymbol{H}}_q \cdot \boldsymbol{A}^n}{\left\|\boldsymbol{H}_q\right\| \times\left\|\boldsymbol{A}^n\right\|} \;,
\label{cos}
\end{array}
\end{equation}
the predicted class is the one with the highest similarity score.
\subsubsection{Retraining}
To further improve the accuracy of the HDC model, retraining can be applied to the AM over several additional iterations on the training set. During retraining, each sample HV is classified. If the prediction is correct, no adjustment is necessary. If the predicted label does not match the true label, the corresponding query HV $\boldsymbol{H}_q$
is subtracted from the misclassified class $\boldsymbol{A}^{\text{miss}}$
and simultaneously added to the real class $\boldsymbol{A}^{\text {real}}$. The process is as follows:
\begin{equation}
\begin{array}{r}
{\boldsymbol{A}^{{\text {miss}}}}={\boldsymbol{A}^{\text {miss}}}-\eta {\boldsymbol{H}_q}\;, \\
{\boldsymbol{A}^{\text {real}}}={\boldsymbol{A}^{\text {real}}}+\eta{\boldsymbol{H}_q}\;,
\end{array}
\label{retraining}
\end{equation}
where $\eta$ is the learning rate.

\subsection{The Framework of FL-HDC-DP}

Our proposed FL-HDC-DP framework combines the computational efficiency of HDC with the privacy protection of DP. 
To protect data privacy during local model transmission, we adopt the zero-concentrated differential privacy (zCDP) mechanism. 
zCDP is particularly well-suited for the iterative nature of FL as it provides a tighter privacy-utility trade-off than standard $(\epsilon,\delta)$-DP. 
In our framework, this guarantee is achieved by adding calibrated Gaussian noise $\mathcal{N}(0,\sigma^2)$ to each device's model update before uploading.
The whole training process, as shown in Algorithm~\ref{alg:fl-hdc-dp}, is summarized as follows:

\textbf{Step 1: Server-side initialization.}
The BS generates bipolar HV benchmarks to ensure consistency in the
position HVs and value HVs across all clients. It also initializes a global AM and broadcasts it to all users, together with the training hyperparameters such as learning rate, DP clipping bound, and the number of local passes per round.

\textbf{Step 2: Client-side local training with DP.}
In the first round, each user encodes all local raw data into sample HVs using \eqref{encoding}, clips them and then aggregates them to built its local AM $\boldsymbol{A}_i$. In subsequent rounds, users receive the global AM $\boldsymbol{A}_g^j$ and perform HDC retraining using local data, updating their local AM to $\boldsymbol{A}_i^{j+1}$ according to \eqref{retraining}, as shown below:
\begin{equation}
\begin{aligned}
& \boldsymbol{A}_{i}^{j+1,\text{miss}}=\boldsymbol{A}_i^{j,{\text{miss}}}-\eta \boldsymbol{H}_{i,q} \\
&\boldsymbol{A}_i^{j+1,\text{real}}=\boldsymbol{A}_i^{j,\text{real}}+\eta \boldsymbol{H}_{i,q}
\end{aligned}
\label{retrain}
\end{equation}
where $\boldsymbol{H}_{i,q}$ is the misclassified query HV of user $i$. 
After encoding or retraining, and before uploading, the user adds Gaussian noise $\boldsymbol{n}_i^{j+1}$ to their local AM $\boldsymbol{A}_i^{j+1}$ to ensure privacy. 
\begin{equation}
 {\boldsymbol{A}}_{i}^{j+1}= {\boldsymbol{A}}_{i}^{j+1} + \boldsymbol{n}_i^{j+1}. 
 \vspace{-0.5em}
\end{equation}
Finally, local AMs are transmitted to the BS. 

\textbf{Step 3: Server aggregation and broadcast.}
Upon receiving the DP-protected local AMs from all users in $\mathcal{U}$, the BS aggregates them by calculating the element-wise average of the class HV to obtain the updated global AM, as given by:
\begin{equation}
{\boldsymbol{A}}_{g}^{j+1}=\frac{\sum_{i=1}^U{\boldsymbol{A}}_{i}^{j+1}}{U}
\vspace{-0.5em}
\end{equation}
The updated global AM $\boldsymbol{A}_{g}^{j+1}$ is then broadcast to all users.

Process 2-3 continues until the global loss function converges or achieves the desired accuracy. 

\subsection{Transmission Model}
After local computing, users upload their DP-added AMs to the BS for aggregation. We adopt frequency division multiple access (FDMA). The achievable uplink transmission rate between user \(i\) and the server with the allocated bandwidth is
\begin{equation}\label{eq:fdma_rate}
r_i \;=\; b_i \log_2\!\left(1+\frac{p_i g_i}{N_0 b_i}\right),
\end{equation}
where \(N_0\) denotes the noise power spectral density; \(b_i\) is the bandwidth allocated to user \(i\); \(p_i\) is the user's transmit power; and \(g_i\) is the end-to-end channel power gain. The per-round bandwidth allocations satisfy $
\sum_{i\in\mathcal{U}} b_i \le B, b_i \ge 0,
$ with \(B\) is the total system bandwidth.
Let \(Q_i\) be the payload size (bits) of the AMs update for user \(i\), which is proportional to the HV dimension. The corresponding transmission time is given by:
\begin{equation}\label{eq:t}
t_i \;=\; \frac{Q_i}{r_i}.
\end{equation}

\subsection{Energy Consumption Model}
In our network, we focus on the energy consumption of each user during the FL training phase, which can be divided into two primary components: a) local computation, including encoding the dataset and HDC model retraining by using its local dataset and the received global AM from server, and b) wireless communication, which involves transmitting the updated local AM.

\subsubsection{Energy Consumption of Computation Model} During the FL training process, in the first round, each user encodes the entire local dataset into HVs and aggregates HVs of the same class to construct the local AM. In the subsequent rounds, the user retrains the HDC model using the encoded HVs together with the global AM received from the server. 

Let \(C_{\rm enc}\), \(C_{\rm agg}\), \(C_{\rm sim}\), and \(C_{\rm up}\) denote the CPU cycles per dimension required for encoding, first‑round aggregation, similarity comparison, and error‑driven update, respectively.  Denote by
$D_i$ is the total number of data sample for user $i$, 
$d$ is HV dimension,
$f_i$ is CPU frequency,
$\gamma$ is the switched‑capacitance coefficient,
$e_i = \frac{O_i}{D_i}$ is inference error rate, and $O_i$ is the number of misclassifications per round.
\begin{itemize}
    \item Round 1 (Encoding and Aggregation): Define the per‑dimension initialization CPU cycles:
$
  C_i^{(\text{init})}
  = C_{\text{enc}} + C_{\text{agg}}
$, so that the latency and energy are given by
\begin{equation}
    \tau_i^{1} = \frac{D_i\,d\,C_i^{(\text{init})}}{f_i}, \quad \forall i \in \mathcal{U},
\label{t1}
\vspace{-0.6em}
\end{equation}
\begin{equation}
    E_i^{1} = \gamma\,D_i\,d\,C_i^{(\text{init})}\,f_i^2.
\label{e1}
\end{equation}
where $D_i\,d\,C_i^{(\text{init})}$ is the total CPU cycles.
 \item Rounds $J\ge2$ (Retraining): Define the per‑dimension retraining CPU cycles:
$
  C_i^{(\text{ret})}
  = C_{\text{sim}}+ e_i\,C_{\text{up}}.
$
Then, the time and energy consumption at user $i$ for retraining the HDC model can be expressed as follows:

\begin{equation}
    \tau_i^{(\text{ret})} = \frac{D_i\,d\,C_i^{(\text{ret})}}{f_i},  \quad \forall i \in \mathcal{U},
\label{tr}
\end{equation}
\begin{equation}
   E_i^{(\text{ret})} = \gamma\,D_i\,d\,C_i^{(\text{ret})}\,f_i^2.
\label{er}
\end{equation}
Eq.\eqref{er} captures the cost of similarity checks plus inference error‑driven updates, with inference error-rate \(e_i\).
\end{itemize}

\subsubsection{Energy Consumption of Transmission Model} After local computation, users transmit their local AM to the server for aggregation via FDMA. Based on Equations \eqref{eq:fdma_rate} and \eqref{eq:t}, the energy consumption for data transmission can be expressed as 
\begin{equation}
    E_i^{(\text{trans})}=t_i\,p_i.
\end{equation}

\subsubsection{Total Energy Consumption}
The total energy consumption of all users participating in FL is given as
\begin{align} 
E=&\underbrace{\sum_{i=1}^{U} \left(E_i^{1}+E_i^{(\text{trans})}\right)}_{\text{First round energy consumption}}+ \underbrace{(J-1)\sum_{i=1}^{U} \left(E_i^{(\text{ret})}+E_i^{(\text{trans})}\right)}_{\text{J-1 rounds energy consumption}},\nonumber\\
=&\sum_{i=1}^U \left(\gamma\,D_i\,d\,C_i^{(\text{init})}f_i^2 + (J-1)\gamma\,D_i\,d\,C_i^{(\text{ret})}\,f_i^2 + J t_i\,p_i\right),
\label{E}
\end{align}

Hereinafter, the total time needed for completing the execution of the FL algorithm is called the completion time. The completion time of each user includes the local computation time and transmission time, based on \eqref{t1} and  \eqref{tr}, the completion time of user $i$ will be 
\begin{align} 
T_i=&\underbrace{\tau_i^{1}+t_i}_{\text{First round time cost}}+ \underbrace{(J-1) \left(\tau_i^{(\text{ret})}+t_i\right)}_{\text{J-1 rounds time cost}},\nonumber \\ 
=&\frac{D_i d \left(C_i^{(\text{init})}+(J-1) C_i^{(\text{ret})}\right)}{f_i} + J \frac{Q_i}{r_i}.
\end{align}

\section{Problem Formulation}\label{optimization}

In this section, we formulate the total energy consumption minimization problem of all users for secure FL-HDC, which jointly considers HDC dimension, power, and computing frequency. The optimization problem is given by:
\vspace{-1em}
\begin{align}
\min_{d,\bm p,\bm f}\quad & \sum_{i=1}^{U}\left(\gamma A_i f_i^2 + J_d\, t_i\, p_i\right) \label{A}\\
\text{s.t.}\quad
& \frac{A_i}{f_i} + J_d\, t_i \le T,\quad \forall i\in\mathcal U \tag{\ref{A}{a}}, \label{Aa}\\
& d_{\min}\le d\le d_{\max} \tag{\ref{A}{b}}, \label{Ab}\\
& 0\le p_i \le P_i^{\max},\quad \forall i\in\mathcal U\tag{\ref{A}{c}}, \label{Ac}\\ 
& 0\le f_i \le f_i^{\max},\quad \forall i\in\mathcal U \tag{\ref{A}{d}}, \label{Ad}
\end{align}
where $\bm p=[p_1,\ldots,p_U]^T$, $\bm f=[f_1,\ldots,f_U]^T$. The per-round compute load is denoted by
$A_i = D_i d C_i$ with $C_i = C_i^{(\mathrm{init})} + (J_d-1) C_i^{(\mathrm{ret})}$. Transmission time is $t_i = \frac{Q_i}{\,b_i \log_2\!\Big(1+\frac{p_i g_i}{N_0 b_i}\Big)}$. $J_d$ represents the number of communication rounds required for a secure HDC-FL model of dimension $d$ to achieve the desired accuracy. Constraint \eqref{Aa} ensures that the combined local execution time and transmission time for each user does not exceed the maximum completion time $T$ for the whole FL algorithm. Since the FL algorithm is performed in parallel and the server updates the global AM only after receiving the uploaded local AMs from all users, this constraint synchronizes the parallel process. The range of the HV dimension is given by \eqref{Ab}. Constraints \eqref{Ac} and \eqref{Ad} limit the transmission power of each user and the maximum local computation frequency, respectively.

\vspace{-0.5em}
\section{Algorithm Design}\label{sec:ALG}

In this section, we develop a low-complexity algorithm that combines enumeration and optimization to solve problem~\eqref{A}. The relationship between $J_d$ and $d$ is first characterized via Monte Carlo simulation. Then, $d$ is enumerated over a finite candidate set $\mathcal{D}$, and for each $d$, the variables $(\boldsymbol{f},\boldsymbol{p})$ are decoupled by enforcing the time constraint in equality form, resulting in a per-user problem with only one variable $f_i$. The optimal dimension $d^{*}$ and corresponding resource allocation $(\boldsymbol{f}^\star,\boldsymbol{p}^\star)$ are finally obtained by minimizing $E$ across all candidates.


\addtolength{\topmargin}{0.03in}
\subsection{Decoupling of $f_i$ and $p_i$}
For each enumerated HDC dimension $d$, the per-user subproblem involves two continuous variables $(f_i,p_i)$.By tightening the time constraint \eqref{Aa} in equality form, the transmit power $p_i$ can be expressed as a function of $f_i$:
\addtolength{\topmargin}{0.03in}
\begin{equation}
p_i = 
\frac{N_0 b_i}{g_i}
\left(
2^{\frac{J_d Q_i}{b_i \left( T - \frac{A_i}{f_i} \right)}} - 1
\right),
\label{p}
\end{equation}
where $T - \frac{A_i}{f_i} > 0$ must hold to ensure feasibility.
Substituting \eqref{p} into energy expression
yields a single-variable optimization problem with respect to $f_i$:
\begin{equation}
 E_i = \gamma A_i f_i^2
+ \left( T - \frac{A_i}{f_i} \right)
\frac{N_0 b_i}{g_i}
\left(
2^{\frac{J_d Q_i}{b_i \left( T - \frac{A_i}{f_i} \right)}} - 1
\right),
\label{Ei}
\end{equation}
and the problem \eqref{A} can be rewritten as:
\begin{align}
\min_{\bm f}\quad & \sum_{i=1}^{U} E_i \label{B}\\
\text{s.t.}\quad
& f_i^{L}\le f_i \le f_i^{\max},\quad \forall i\in\mathcal U \tag{\ref{B}{a}}. \label{Ba}
\end{align}
Where the $f_i^{L}$ is determined by the $P_i^{\text{max}}$ given by
\begin{equation}
   f_i^{\mathrm{L}} = 
\frac{A_i}{
T - 
\frac{J_d Q_i}{
b_i \log_2\!\left( 1 + \frac{g_i P_i^{\max}}{N_0 b_i} \right)}}. 
\end{equation}

Since $A_i$, $b_i$, $g_i$, and $Q_i$ are constants for a given $d$, problem \eqref{B} is separable across users, and each subproblem $E_i$ is one-dimensional. 
The objective $E_i$ consists of two convex terms: 
a quadratic local-computation term $\gamma A_i f_i^2$ and a monotonically increasing convex transmission-energy term induced by the exponential function $2^{\frac{J_d Q_i}{b_i(T - A_i/f_i)}}$. 
Both are convex on the feasible interval $[f_i^{L}, f_i^{\max}]$, and hence $E_i$ is strictly convex and admits a unique minimizer.

Therefore, the optimal $f_i$ for under dimension $d$ is obtained via a bounded one-dimensional search such as Brent's method:
\begin{equation}
    f_i^\star(d)
    = \arg\min_{f_i^{L} \le f_i \le f_i^{\max}} E_i,
    \label{eq:fstar}
\end{equation}
which guarantees convergence to the global optimum. 
Then, the corresponding transmit power is recovered by substituting \eqref{eq:fstar} into \eqref{p} as $
    p_i^\star(d) = p_i\bigl(f_i^\star(d)\bigr).
$
The total energy for a given dimension $d$ is
\begin{equation}
    E(d) = \sum_{i=1}^{U} E_i\bigl(f_i^\star(d)\bigr).
    \label{eq:Ed}
\end{equation}

\subsection{Outer Enumeration over HDC Dimension}
The overall optimization is accomplished by enumerating all candidate HDC dimensions in the finite set $\mathcal{D}$ and selecting the one that minimizes the total energy in \eqref{eq:Ed}:
\begin{equation}
    d^\star = \arg\min_{d \in \mathcal{D}} E(d),
    \label{eq:dstar}
\end{equation}
and the corresponding resource allocation
$\bigl\{f_i^\star(d^\star), p_i^\star(d^\star)\bigr\}$ 
is adopted as the final solution.

\subsection{Complexity Analysis}
Let $D=|\mathcal{D}|$ denote the number of candidate HDC dimensions. For each $d \in \mathcal{D}$, the optimization in \eqref{eq:fstar} is performed independently for all users through a one-dimensional search. 
If the search requires at most $L$ function evaluations to reach convergence, the overall computational complexity is
$
    \mathcal{O}\bigl(DUL\bigr).
$
Since $D$ and $L$ are small constants, the proposed enumeration-and-optimization framework scales linearly with the number of users and is well suited for FL deployment at resource-constrained edge devices.

\section{Simulation Results and Analysis}\label{simulation}
We consider a secure FL-HDC system in which $K=50$ users are independently and uniformly distributed over a circular area of radius $500$\,m, with the BS located at the center. Experiments are conducted on the MNIST dataset. All system and computation parameters are summarized in Table~\ref{tab:sim_params}. Since it is challenging to mathematically express the relationship between privacy, dimension, and convergence rounds required to achieve the same accuracy, we employ Monte Carlo simulations to obtain the relationship between dimension and convergence rounds for an accuracy of 88\% under $(\varepsilon,\delta)=(25,10^{-5})$, as shown in Table \ref{tab:convergence_dimension}.

\begin{table}[t]
\centering
\caption{Simulation parameters}
\label{tab:sim_params}
\renewcommand{\arraystretch}{1.1}
\begin{tabular}{l l}
\hline
\textbf{Parameters} & \textbf{Value} \\
\hline
$U$: number of users & $50$ \\
$C_{\mathrm{enc}}$: encoding cycles per dimension & $28*28*2$  \\
$C_{\mathrm{agg}}$:  1st-round aggregation cycles per dimension& $28*28*2$ \\
$C_{\mathrm{sim}}$: similarity evaluation cycles per dimension & $10*10$  \\
$C_{\mathrm{up}}$: error-driven update cycles per dimension& $8$  \\
$e_i$: inference error ratio for user $i$ & $0.4$ \\
$d$ : HV dimension range & 3000:1000:10000\\
$Q_i$: upload size of user $i$ & 10*$d$\\
$S_i$: number of local data for user $i$ & 1200\\
$\alpha$: learning rate of HDC model & $ 1$\\
$f_i^{\max}$: maximum CPU frequency of user $i$  & $2.3$ GHz \\
$\gamma$: switched-capacitance coefficient & $1\times10^{-28}$ \\
$B$: total bandwidth & $1$ MHz \\
$P^{\text{max}}$: maximum transmit power & $0.1-1$ W \\
$n_0$: noise power spectral density & $-174$ dBm/Hz \\
\hline
\end{tabular}
\vspace{-4pt}
\end{table}

\begin{table}[t] \centering \caption{Iterations to Converge under Accuracy $=88\%$ and Privacy Threshold $\epsilon=25$} \label{tab:convergence_dimension}
\setlength{\tabcolsep}{4pt} 
\begin{tabular}{c|cccccccc} \hline Dimension $d$ & 3000 & 4000 & 5000 & 6000 & 7000 & 8000 & 9000 & 10000 \\ \hline Iterations $J_d$ & 39 & 20 & 18 & 17 & 16 & 16 & 14 & 14 \\ \hline \end{tabular} 
\vspace{-1.9em}
\end{table}

\begin{figure}[t]
    \centering
\includegraphics[width=0.86\linewidth]{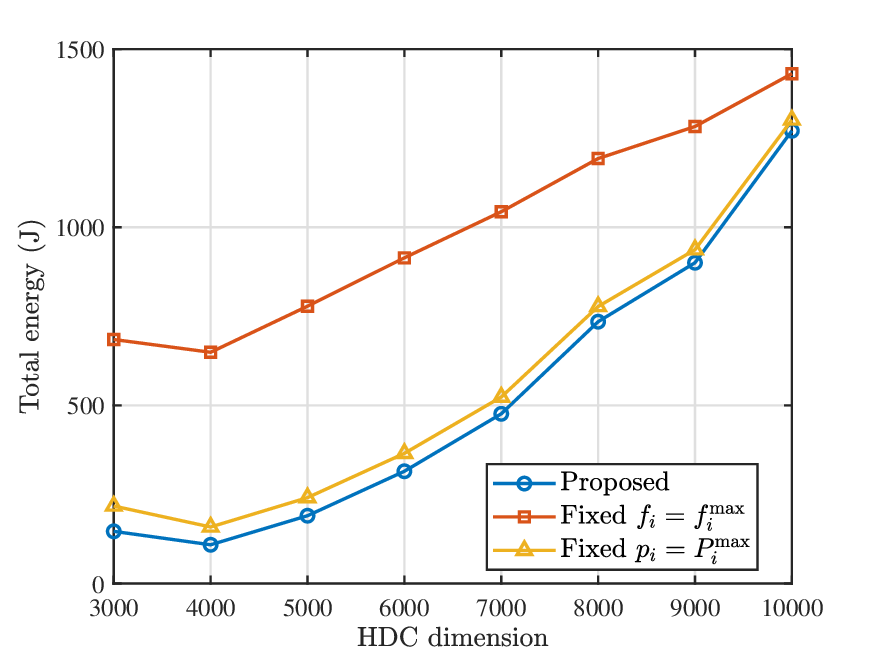}
\vspace{-0.4em}
    \caption{Total energy versus HDC dimension with $T=30s$.}
    \label{fig:1}
    \vspace{-1.5em}
\end{figure}

Fig. \ref{fig:1} illustrates the total energy consumption with respect to the HDC dimension when the total time is fixed at $T=30s$. The results indicate that the total energy does not vary monotonically with $d$; instead, the curve reaches its minimum at a moderate dimension ($d=4000$). According to Table~\ref{tab:convergence_dimension}, when $d$ increases from 3000 to 4000, the number of global rounds required to achieve the target accuracy decreases by 19. Although each round incurs higher local computation and transmission costs due to the additional 1000 dimensions, this extra overhead is offset by the cumulative energy savings from skipping 19 rounds, resulting in lower total energy consumption. When the dimension further increases to 5000, the number of required rounds decreases by only about two more, but the per-round energy increase outweighs the savings from fewer rounds, leading to a rise in total energy. Compared with the two baseline schemes, the proposed method achieves the lowest overall energy across all dimension settings, reducing total energy consumption by approximately 83.3\% and 31.5\% relative to the fixed-frequency and fixed-power schemes, respectively.
\begin{figure}[t]
    \centering    \includegraphics[width=0.86\linewidth]{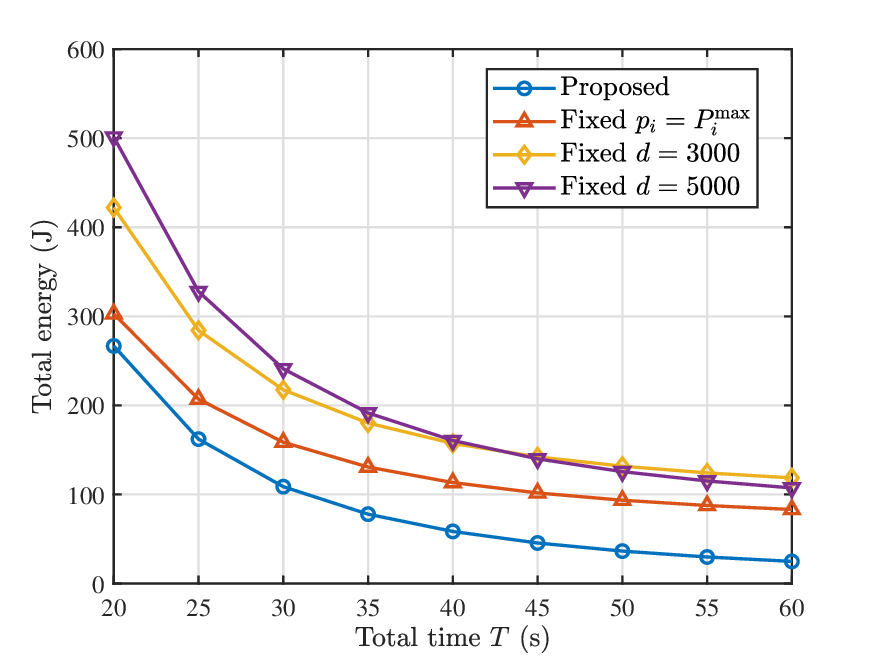}
    \vspace{-0.5em}
    \caption{Total energy consumption versus total time.}
    \label{fig:2}
    \vspace{-1.5em}
\end{figure}

Fig.~\ref{fig:2} compares the total energy consumption of the proposed scheme with three baseline schemes under different total time constraints. 
As $T$ increases, the total energy consumption of all schemes decreases, since a longer execution time allows lower transmit power and CPU frequency while still satisfying the time constraint. 
When $T$ is small, communication becomes the dominant bottleneck, forcing users to transmit at high power and operate with large CPU frequencies $f_i$, which leads to high total energy consumption. 
As $T$ relaxes, more time is allocated for local computation, allowing $f_i$ to decrease and thus reducing the total energy. 
In addition, the proposed method consistently achieves the lowest total energy across all time settings, reducing total energy consumption by approximately 
54.9\%, 50.0\%, and 31.5\% compared with the $d = 3000$, $d = 5000$, and fixed-power baselines, respectively. 
The results also reveal that the impact of the HDC dimension is substantial. The optimal dimension is $d = 4000$, and it is clearly observed that the total energy at $d = 3000$ and $d = 5000$ is much higher than that at $d = 4000$, which is consistent with the conclusion drawn from Fig.~\ref{fig:1}. 

\section{Conclusion}\label{con}

In this paper, we proposed a secure FL framework based on HDC and DP for resource-constrained edge devices in wireless networks. We formulated a joint optimization problem of the HDC dimension, transmit power, and CPU frequency to minimize total energy consumption, which is the first study to consider the impact of the HDC dimension on energy optimization. To efficiently solve this problem, we designed a two-stage enumeration and optimization framework that first characterizes the relationship between HDC dimension and convergence behavior, and then optimizes resource allocation under latency constraints. Simulation results illustrate that the proposed scheme significantly reduces total energy compared with baseline methods, with dimension selection playing the most critical role among all optimization variables.
\section*{Acknowledgment}
This work was supported in part by the U.S. National Science Foundation under Grant SaTC-2350076.



%


\ifCLASSOPTIONcaptionsoff
  \newpage
\fi



%

%

\bibliographystyle{IEEEtran}
\bibliography{ref}
\end{document}